\documentclass{llncs}

\usepackage{dsfont}
\usepackage{amsmath}
\usepackage{authblk}
\usepackage[utf8]{inputenc}
\usepackage[OT1]{fontenc}
\newcommand{\G}{\mathds{G}}
\newcommand{\Z}{\mathds{Z}}

\newcommand{\set}[1]{\{#1\}}
\newcommand{\abs}[1]{\left|#1\right|}

\usepackage[printonlyused]{acronym}

\begin{document}

\title{Oblivious Lookup Tables}

\author{Stefan Rass$^1$\and Peter Schartner$^1$\and Markus Wamser$^2$}
\institute{$^1$ Institute of Applied Informatics, Universit\"{a}t Klagenfurt,
Universit\"{a}tsstrasse 65-67, 9020 Klagenfurt, Austria, \{stefan.rass,
peter.schartner\}@aau.at,\\
$^2$ Institute for Security in Information Technology, Technical University
of Munich, Arcisstraße 21, 80333 Munich, Germany, wamser@tum.de}

%
%
%
%

\maketitle              
\abstract{We consider the following question: given a group-homomorphic
public-key encryption $E$, a ciphertext $c=E(x,pk)$ hiding a value $x$ using
a key $pk$, and a "suitable" description of a function $f$, can we evaluate
$E(f(x), pk)$ without decrypting $c$? We call this an \emph{oblivious lookup
table} and show the existence of such a primitive. To this end, we describe a
concrete construction, discuss its security and relations to other
cryptographic primitives, and point out directions of future investigations
towards generalizations.}

\noindent\textbf{Keywords.} homomorphic encryption; private function
evaluation; secure function evaluation


%

\section{Introduction and Concept}
This short note is about the following setting: let $f:X\to Y$ be a mapping
between finite sets. Assume that the sizes of $X$ and $Y$ are sufficiently
small to permit a specification of $f$ via a lookup-table. Let $E(m,\kappa)$
denote a group-homomorphic encryption of a message $m$ under a key $\kappa$,
where $E$ can be symmetric or asymmetric. Let $E$ be group-homomorphic in the
sense that $E(m_1\cdot m_2,\kappa)=E(m_1,\kappa)\cdot E(m_2,\kappa)$, where
$\cdot$ denotes the respective group operations within the plain- and
ciphertext space.

In this setting, we consider the following question: given $E$ and an
encrypted value $c=E(x,\kappa)$, can we compute $E(f(x),\kappa)$ without
decrypting $c$? We call any such implementation of $f$ an \emph{oblivious
lookup table}, as it shall effectively hide the evaluation of $f$, or
equivalently, evaluate $f$ only on ciphertexts by virtue of conventional
homomorphic encryption.

Becoming more specifically, let $p=2q+1$ be a safe prime (i.e., $q$ is a
prime too), and let $\G\subset\Z_p$ denote the $q$-order subgroup generated
by some element $g\in\Z_p$. We first describe the lookup technique in plain
form, and subsequently wrap the encryption around the necessary operations.

Let $X=\set{x_1,\ldots,x_n}\subseteq\G$ be an enumeration of (distinct)
values to be looked up. To each such element $x_i$, we associate a vector
$\vec v_i = (x_i^k)_{k=0}^{n-1}=(1,x_i,x_i^2,\ldots,x_i^{n-1})$. Notice that
$x_i\neq x_j$ whenever $i\neq j$ implies that the vectors $\vec
v_1,\ldots,\vec v_n$ are all linearly independent, as they essentially form
the rows of a Vandermonde-matrix $\vec V$. Without loss of generality, let us
assume $\abs{X}=n=\abs{Y}$, say, by allowing multiple occurrences of the same
element in $Y$ in case that $f$ is not injective. Under this convention, let
the (not necessarily pairwise distinct) elements of $Y$ be enumerated as
$Y=\set{y_1,\ldots,y_n}$.

We will construct the value of $f(x_i)$ by a scalar product of $\vec v_i$
with a vector-representation of the lookup table. That is, the lookup table
itself is a vector $\vec \ell$ with the property that $\vec v_i^T\cdot\vec
\ell=f(x_i)$ for all $i=1,2,\ldots,n$. To this end, let us choose an
arbitrary but invertible matrix $\vec U\in\G^{n\times n}$ with columns $\vec
u_1,\ldots,\vec u_n$. Define the lookup table as $\vec \ell:=\vec
U\cdot\vec\alpha$ for some (yet to be determined) vector $\vec
\alpha=(\alpha_1,\ldots,\alpha_n)$. Now, let us look at the scalar product of
$\vec v_i$ with $\vec U\cdot \vec\alpha$ to yield $f(x_i)\in\Z_p$. This
results in a linear equation $\alpha_1(\vec v_i^T\cdot \vec
u_1)+\alpha_2(\vec v_i^T\cdot \vec u_2)+\cdots+\alpha_n(\vec v_i^T\cdot\vec
u_n)=f(x_i)$. Establishing this condition for all $i=1,2,\ldots,n$, we end up
observing that, to find $\vec \alpha$, we need to solve the linear system
$(\vec V\cdot\vec U)\cdot\vec\alpha=(f(x_1),\ldots,f(x_n))^T$ for $\vec
\alpha$. The coefficient matrix $\vec V\cdot\vec U$ is invertible by
construction, and hence we can easily lookup values $f(x_i)$ by computing
$f(x_i)=\vec v_i^T\cdot\vec\ell$, taking $O(n)$ multiplications and
additions.

Now, let us see if we can equivalently do all the necessary steps when the
pre-image is encrypted. For that matter, we take an element-wise commitment
to the $\vec v_i$ from above to represent $x_i$. That is, the value $x_i$ now
comes committed and encrypted as $\tilde E(x_i,\kappa):=(E(1,\kappa)$,
$E(g^{x_i},\kappa)$, $E(g^{x_i^2},\kappa)$,\ldots,
$E(g^{x_i^{n-1}},\kappa))=(c_1,\ldots,c_n)$, so that the matrix of exponents
remains $\vec V=(v_{ij})_{i,j=1}^n$ with $v_{ij}=x_i^{j-1}$ and as such
invertible. Since the order of $\G$ is a prime, we can -- in a setup phase
where the exponents are known -- straightforwardly work out the values
$\vec\alpha$ and the lookup table $\vec \ell=(\ell_1,\ldots,\ell_n)$, which
is supplied in plain (not encrypted) form to the instance that seeks to
evaluate $f$.

To evaluate $f$, let the encrypted input value $x_i$ be given as $\tilde
E(x_i,\kappa)$. Then, we can compute the lookup value $E(f(x_i),\kappa)$ as

\begin{align}
    \prod_{k=1}^n c_k^{\ell_k} &= \prod_{k=1}^n E(g^{x_i^{k-1}},\kappa)^{\ell_k}
    = \prod_{k=1}^n E(g^{v_{ik}},\kappa)^{\alpha_1 u_{k1} + \alpha_2 u_{k2} + \ldots \alpha_n u_{kn}}\nonumber\\
    &=\prod_{k=1}^n E(g^{\alpha_1 v_{ik} u_{k1}+\alpha_2 v_{ik}u_{k2}+\ldots+\alpha_n v_{ik}u_{kn}},\kappa) = E(g^{f(x_i)},\kappa)\label{eqn:lookup}.
\end{align}
The last equality is instantly obtained by writing out the exponents for
$k=1,2,\ldots,n$ and rearranging terms properly when summing up.

A final remark is judicious here: the formula yields only a single value
based on an input vector. To properly implement the lookup to be
\emph{repeatable}, i.e., to model iterations like $f(f(\cdots f(x)\cdots))$
or generally functions $f:X\to X$, we need to look up all the elements of the
output vector via separate tables. So, the overall lookup table is no longer
a $n$-dimensional vector, but an $(n\times n)$-matrix $\vec L=(\vec
\ell_1,\ldots,\vec \ell_n)$. The $j$-th such lookup table $\vec \ell_j$ must
then be designed to return $y^{j-1}$, whenever the input value $x$ is
represented by a sequence $1,x,x^2,\ldots,x^{n-1}$ in the exponents. That is,
the mapping $f(x)=y$, acting on $x$ being represented by encrypted values
$1,g^x,g^{x^2},\ldots,g^{x^{n-1}}$, requires $n$ lookups that successively
yield $1,g^y,g^{y^2},\ldots,g^{y^{n-1}}$, each of which by \eqref{eqn:lookup}
requires $O(n)$ exponentiations and multiplications. So, the total cost of an
oblivious lookup comes to $O(n^2)$ exponentiations (subsuming multiplications
as the cheaper operation here).

Considering security, each lookup table is indeed available in plaintext, but
since it is independent of a particular input and the input and output values
remain encrypted at all times, knowledge of $\vec L$ cannot release any
information about the secret $x$ being transferred into the secret result
$f(x)$. Probabilistic encryptions like ElGamal an offer the additional appeal
of enforced re-randomization of the resulting ciphertexts. That is, if a
distrusted third party does several lookups, it nevertheless cannot recognize
any results as being identical to previous ones.

\section{Related Work}

This work closely relates to \ac{PFE}, which provides a system where the
function-to-be-evaluated $f$ \emph{and} the inputs are private and the
evaluator learns nothing about either aside from the (encrypted) results of
the evaluation of the function on the inputs. This can be realized using
\ac{SFE} over a universal circuit
(\cite{KolesnikovSchneider2008,Valiant1976}), to which $f$ has to be
converted first. Another approach is to use a (non-universal) circuit
representation of $f$ and employ a \ac{FHE} scheme
\cite{Gentry2009,Silverberg2013}. However, all mentioned approaches carry
complexities that are too high for practical applications. Conceptually
closest to our ideas seem to be \cite{KatzMalka2011} and
\cite{MohasselSadeghian2013}, both based on singly homomorphic encryption.
The former realises \ac{PFE} in a strict two-party setting with one party
providing the function and the other providing the inputs. Evaluation is done
through a common virtual machine. The latter is based on a framework that
splits the task into \ac{CTH} and \ac{PGE} which together enable \ac{PFE}
with linear complexity in all standard settings. However, both \ac{PFE}
protocols require an interactive setting while we are aiming for the
non-interactive setting. The security implications tied to our simple scheme
when being lifted to two-operand functions (if that it possible at all) are,
however, far from clear and probably intricate (cf. \cite{Boneh2011}) and
will be discussed along the research sketched in this abstract.

\section{Open Issues}
A yet open issue is a proper formalization of security for oblivious lookup
tables. Intuitively, the attacker should be unable to learn anything
meaningful from the lookup table as such, since this is nothing but a bunch
of ciphertexts and hence indistinguishable from self-made cryptograms,
provided that the encryption is semantically secure. However, a full-fledged
formal argument and definition of security is subject of future
considerations. Also, the idea does not obviously generalize to functions of
multiple inputs, which poses another interesting question for future
research.

\bibliographystyle{plain}
\bibliography{cecc15}

\section*{Acronyms}
\begin{acronym}
\acro{CTH}{Circuit Topology Hiding}
\acro{FHE}{Fully Homomorphic Encryption}
\acro{PFE}{Private Function Evaluation}
\acro{PGE}{Private Gate Evaluation}
\acro{SFE}{Secure Function Evaluation}
\end{acronym}

\end{document}